\documentclass[fleqn,usenatbib]{mnras}

\usepackage{newtxtext,newtxmath}
\usepackage[T1]{fontenc}

\DeclareRobustCommand{\VAN}[3]{#2}
\let\VANthebibliography\thebibliography
\def\thebibliography{\DeclareRobustCommand{\VAN}[3]{##3}\VANthebibliography}

\usepackage{graphicx}	% Including figure files
\usepackage{amsmath}	% Advanced maths commands

\title[The magnetic mayhem in Abell 2199]{The magnetic mayhem in Abell 2199: discovery of synchrotron threads and homogeneous diffuse radio lobes}

\author[R. Timmerman et al.]{
R. Timmerman,$^{1,2}$\thanks{E-mail: roland.timmerman@durham.ac.uk (RT)}
L. Rudnick,$^{3}$
A. Botteon,$^{4}$
G. Brunetti,$^{4}$
and R. Kale,$^{5}$
\\
$^{1}$Centre for Extragalactic Astronomy, Department of Physics, Durham University, Durham, DH1 3LE, UK\\
$^{2}$Institute for Computational Cosmology, Department of Physics, Durham University, South Road, Durham, DH1 3LE, UK\\
$^{3}$Minnesota Institute for Astrophysics, University of Minnesota, 116 Church St. SE, Minneapolis, MN 55455, USA\\
$^{4}$INAF - Instituto di Radioastronomia, via P. Gobetti 101, 40129 Bologna, Italy\\
$^{5}$National Centre for Radio Astrophysics, Tata Institute of Fundamental Research, S. P. Pune University Campus, Ganeshkhind, Pune 411007, India\\
}

\date{Accepted 2026 June 25. Received 2026 June 25; in original form 2026 May 31}

\pubyear{\the\year{}}

\begin{document}
\label{firstpage}
\pagerange{\pageref{firstpage}--\pageref{lastpage}}
\maketitle

\begin{abstract}
Sensitive low-frequency radio observations have started uncovering examples of synchrotron-emitting threads, isolated from the rest of radio emission in galaxy clusters. As the bridge of radio emission previously detected between the radio lobes of 3C\,338 in Abell\,2199 is a candidate of such a structure, we observed this galaxy cluster using the International LOFAR Telescope. These observations revealed the presence of multiple narrow isolated synchrotron threads in 3C\,338: east, west and north of the AGN and its radio lobes. Chandra X-ray observations show that these structures most likely do not reside within cavities in the intracluster medium (ICM), and are therefore considered to be distinct structures from the radio lobes. Non-detections in 1.5~GHz Very Large Array observations imply that the spectral index of these newly-discovered isolated threads is likely \(\alpha_{1500}^{144} < -3.0\) or steeper. We consider these isolated synchrotron threads to most likely display examples of magnetic threads within the ICM that have captured synchrotron-emitting plasma, as has recently been proposed.
Furthermore, our observations reveal the radio lobes to show an almost perfectly uniform spectral index, unlike what would be expected if substantial age differences are present in the radio lobes according to standard spectral ageing models. We find that the relativistic plasma in 3C\,338 is consistent with a homogeneous cosmic ray electron population, with the spectral variations dependent on the local magnetic field strength. Finally, we explore the various models that could explain this trend in the radio lobes.
\end{abstract}


%%%%%%%%%%%%%%%%%%%%%%%%%%%%%%%%%%%%%%%%%%%%%%%%%%

%%%%%%%%%%%%%%%%% BODY OF PAPER %%%%%%%%%%%%%%%%%%

\section{Introduction}

Recent developments of sensitive low-frequency radio observatories have led to the unexpected detection of isolated synchrotron-emitting threads associated with active galactic nuclei (AGN) in galaxy clusters. Most notably, \citet{ramatsoku20} reported the discovery of such collimated synthrotron threads between the radio lobes of ESO\,137-006 using MeerKAT observations \citep{jonas16} at 1000 and 1400~MHz. Briefly thereafter, \citet{rudnick22} detected synchrotron-emitting threads extending away from the Northern radio lobe in 3C40B using both MeerKAT and the LOw Frequency ARray \citep[LOFAR;][]{haarlem13}.

Currently, our knowledge and understanding of the physical nature of these structures remains very limited due to the low number of detections in observations. The two aforementioned examples both show synchrotron-emitting threads extending on the order of 100~kpc with a diameter on the order of 1~kpc. Their radio spectra at around 1~GHz feature spectral indices of \(\alpha \approx -2\), where we define our spectral indices (\(\alpha\)) according to \(S_\nu \propto \nu^\alpha\), with \(S_\nu\) being the surface brightness and \(\nu\) the frequency. However, they also fundamentally differ in structure, as in ESO\,137-006 one of the threads connects the two radio lobes, and in 3C40B the threads are seen to be extending away from one of the radio lobes into the cluster environment. The synchrotron threads in both of these examples display an unforeseen behavior of the radio lobes. 
Simulations suggest that synchrotron threads like in ESO\,137-006 could form due to the interaction between the relativistic plasma ejected by the AGN and the magnetic field structures within a galaxy cluster \citep[e.g.,][]{lalakos25}. This makes synchrotron threads potentially unique observational probes of the magnetic field structures present in galaxy clusters, as well as a test of the evolution of relativistic plasmas in these environments \citep[e.g.,][]{churazov26}.

Thanks to breakthroughs in the calibration of long-baseline LOFAR observations \citep{morabito22, sweijen22}, we now have access to sub-arcsecond resolution radio observations at 144~MHz. This is more than an order-of-magnitude improvement compared to the angular resolution of 6~arcseconds attainable using only the Dutch part of LOFAR, such as used by the LOFAR Two-Metre Sky Survey \citep[LoTSS,][]{shimwell17,shimwell19,shimwell22,shimwell26}. Because of the excellent combination of high angular resolution and a low observing frequency, these observations are ideally suited to study steep-spectrum narrow threads. For this reason, we targeted the radio source 3C\,338 within the galaxy cluster Abell\,2199 with the International LOFAR Telescope (ILT), as indications of a narrow thread connecting the two radio lobes in this system were previously reported by \citet{burns1983}. Here, we report on the discovery of multiple previously undetected isolated synchrotron threads in 3C\,338, as well as on the radio lobes and the bridge between them directly south of the AGN core.

We adopt a \(\Lambda\)CDM cosmology with a Hubble parameter of \(H_0\ =\ 70~\mathrm{km}\ \mathrm{s}^{-1}\ \mathrm{Mpc}^{-1}\), a matter density parameter of \(\Omega_m\ =\ 0.3\), and a dark energy density parameter of \(\Omega_\Lambda\ =\ 0.7\). With these parameters and at the redshift of 3C\,338 \citep[\(z=0.031\);][]{minkowski61}, 1\arcsec = 0.620~kpc. All uncertainties stated denote the 68.3\%~=1\(\sigma\) confidence interval.

\section{Observations and data reduction}

\begin{figure*}
    \centering
    \includegraphics[width=\textwidth]{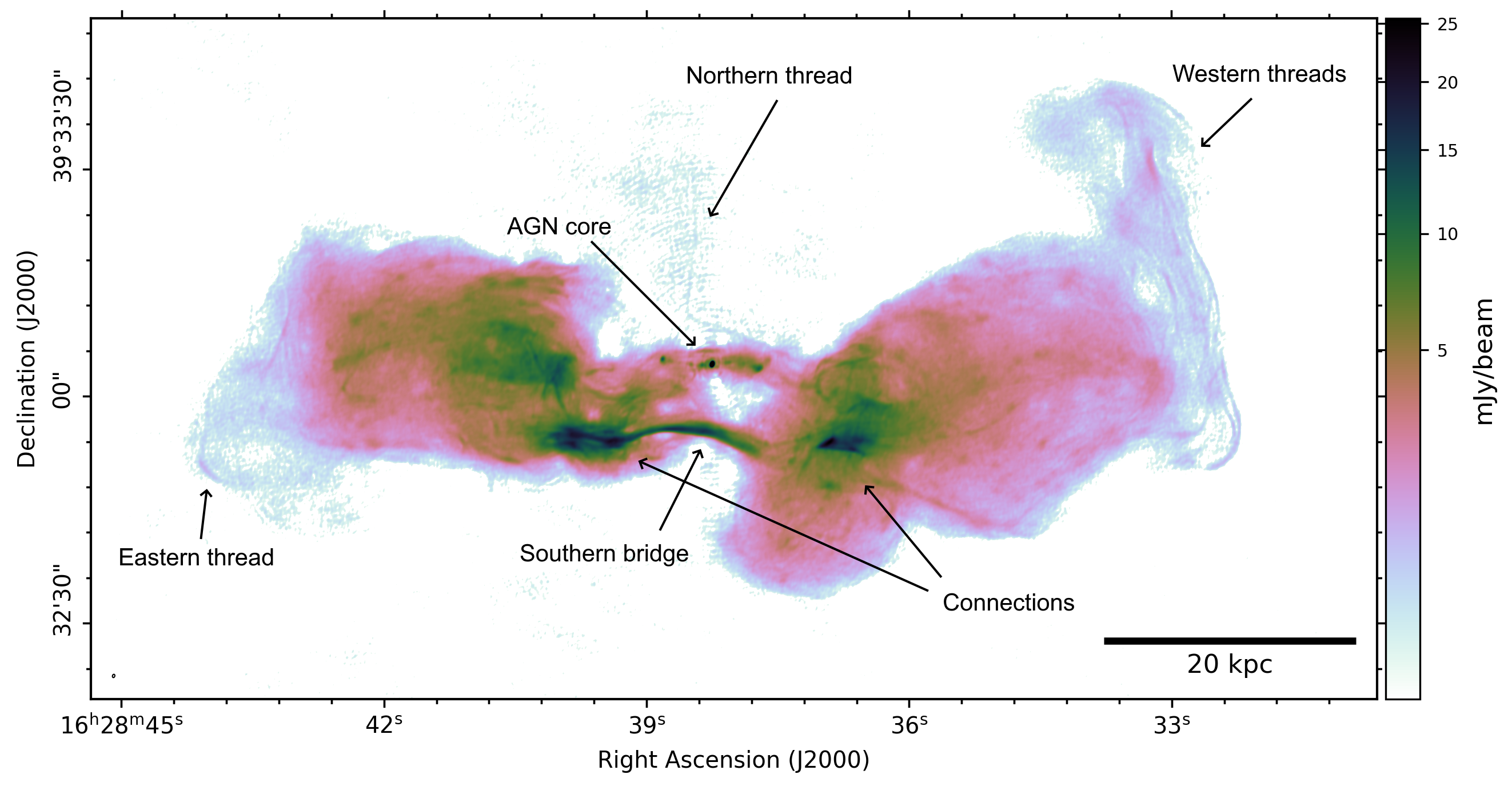}
    \caption{LOFAR-VLBI image of 3C\,338 at an observing frequency of 144~MHz. The color scale goes from three times the rms noise level (\(\sigma_\mathrm{rms}=70\)~\textmu Jy/beam) to the peak brightness of the radio lobes. The AGN core extends beyond the color scale and reaches a peak brightness of 189 mJy/beam. The scale bar in the bottom right corner measures 20~kpc at the redshift of 3C\,338. The synthesized beam size (\(0.442^{\prime\prime}\times0.309^{\prime\prime}, \mathrm{PA}=-16.7^\circ\)) is indicated by the ellipse in the bottom-left corner.}
    \label{fig:lofar_map}
\end{figure*}

\subsection{LOFAR}

The radio source in Abell\,2199 (3C\,338) was observed on 15 November 2020 with LOFAR for a total of 8~hours between 120 and 168~MHz (Project code: LC14\_019, PI: R. Timmerman). The standard calibrators 3C\,147 and 3C\,295 were observed for 10~minutes before and after 3C\,338, respectively, for the calibration of the polarization alignment, the clock offsets and the bandpass. The calibration was performed using the standard LOFAR calibration \textsc{Prefactor} software package \citep[now \textsc{LINC};][]{vanweeren16, williams16, gasperin19}. First, radio-frequency interference (RFI) was flagged from the data using \textsc{AOFlagger} \citep{offringa13, offringa15} to exclude affected visibilities from further steps. Next, the visibilities were calibrated against a source model to derive corrections for the polarization alignment, the Faraday rotation introduced within Earth's ionosphere, the bandpass, and clock differences between individual LOFAR stations. With the previous calibration solutions applied to the data, the phases of the target visibilities from the Dutch stations were calibrated against a TIFR Giant Metrewave Radio Telescope Sky Survey \citep[TGSS,][]{intema17} sky model.

Next, the calibration was extended to the international LOFAR stations using the LOFAR-VLBI pipeline \citep{morabito22}, in which \textsc{facetselfcal} was employed to perform dispersive delay calibration \citep{vanweeren21}. With the visibilities phase-shifted towards 3C\,338, the core stations of LOFAR were first phased up into a single super station to narrow down the field of view and reduce interference from surrounding radio sources. The data were then averaged down to 8-second long integrations and frequency channels of 97~kHz. Finally, the dispersive phase and residual gain corrections were calculated by self-calibrating on 3C\,338, starting with a point-source model to represent the bright AGN core.

\subsection{Very Large Array}

For spectral index mapping, radio observations at higher frequencies are required. Very Large Array (VLA) observations taken using the L and S bands were obtained in calibrated form from the VLA archive \citep{kent18}. These were then combined per receiver band and self-calibrated together using the Common Astronomy Software Application \citep[\textsc{CASA};][]{mcmullin07}. The self-calibration included determining both phase and amplitude corrections to obtain the final calibrated data sets. The data sets obtained from the VLA archive are summarized in Table~\ref{tab:VLA}.

\begin{table}
    \caption{Summary of the VLA observations obtained of 3C\,338.}
    \centering\small
    \begin{tabular}{lllll}\hline\hline
        Receiver & Config. & Project code & Date & Duration\\\hline
        L band & B & 19A-132 & 28 April 2019 & 40 min.\\
        L band & B & 19A-132 & 2 May 2019 & 40 min.\\
        L band & B & 19A-132 & 23 June 2019 & 31 min.\\
        L band & C & 18B-159 & 2 Dec. 2018 & 39 min.\\\hline
        S band & B & 19A-132 & 28 April 2019 & 40 min.\\
        S band & B & 19A-132 & 2 May 2019 & 39 min.\\
        S band & B & 19A-132 & 23 June 2019 & 31 min.\\
        S band & C & 18B-159 & 2 Dec. 2018 & 40 min.\\\hline\hline
    \end{tabular}
    \label{tab:VLA}
\end{table}

\begin{figure*}
    \centering
    \includegraphics[width=\textwidth]{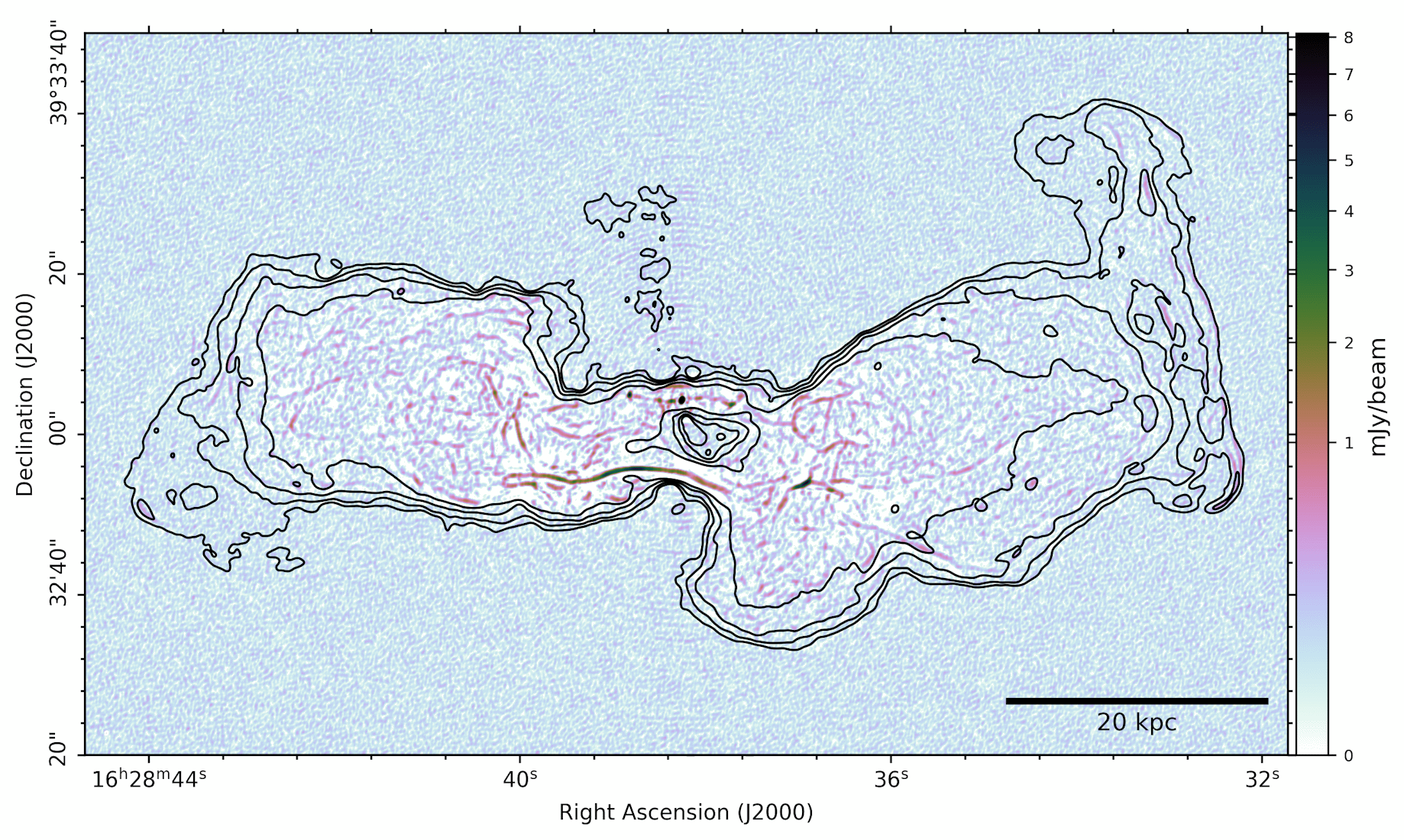}
    \caption{Filtered version of the LOFAR-VLBI image of 3C\,338 from Fig.~\ref{fig:lofar_map} obtained using the filtering procedure described by \citet{rudnick02}. This procedure separates small-scale features from large-scale features in an image using ``top-hat'' filters. Here, the small-scale features extracted from the image are shown. We applied a filtering scale of 9 pixels, which corresponds to \(0.675^{\prime\prime}=0.42\ \mathrm{kpc}\). The black contours indicate the smoothed total 144~MHz intensity from Fig.~\ref{fig:lofar_map}.}
    \label{fig:lofar_filtered}
\end{figure*}

\subsection{Chandra}

We retrieved from the Chandra Data Archive all available observations of Abell\,2199. These consist of two ACIS-S observations (ObsIDs: 497, 498), totalling 40~ks of observing time, and four ACIS-I observations (ObsIDs: 10748, 10803, 10804, 10805), totalling 120~ks of observing time. Data were reprocessed from the level-1 using CIAO v4.12. We extracted light curves in the 0.5-7.0~keV band to filter out periods characterized by high-background levels. The final total cleaned exposure time is 156~ks. The six ObsIDs were reprojected to the same tangent point and merged to produce a combined exposure-corrected image in the 0.5-2.0~keV band with a pixel-scale of 0.492~arcseconds. When combining the datasets, we considered only the chips I0-I3 of the ACIS-I observations and the S3 chip of the ACIS-S observations.

\section{Results}

\subsection{Low-frequency imaging}

Following the data reduction and calibration, the LOFAR data at 144~MHz were imaged using \textsc{WSClean} \citep{offringa14, offringa17} using multiscale cleaning. The resulting image has a resolution of \(0.44^{\prime\prime}\times0.31^{\prime\prime}\) with an off-source rms noise level of 65~\textmu Jy/beam, and is shown in Fig.~\ref{fig:lofar_map}.

In addition to the extended radio lobes and the compact bright AGN core of 3C\,338 at the center of the cluster, a set of features is revealed in the LOFAR map. First of all, the bridge of radio emission south of the AGN core first reported by \citet{burns1983} is clearly detected. Consistent with previous observations, the bridge appears to connect to both radio lobes and lacks a direct connection to the AGN core. Furthermore, the high-resolution LOFAR map shows a narrow thread extending beyond the eastern lobe, apparently connected to the lobe at both ends of the thread. There is also a thread north of the AGN core which appears to connect to the inner edge of the western radio lobe, directed back towards the AGN. Most prominently, a bundle of threads on the far side of the western lobe is revealed. This bundle appears to contain not simply a discrete set of narrow threads, but rather a continuum of threaded structure that likely proceeds below the resolution limit of the observation. For the most part, these threads appear to drape around the surface of the radio lobe, although some appear to attach to the lobe on the southern end, and a ``hammerhead''-like structure is visible at the northern end. These threads also show up in the low-resolution LOFAR map published by \citet{birzan20}, though this map did not provide the angular resolution to resolve the threads within this region. We note that threaded structures are visible internal to the radio lobes as well, primarily closer to the center of the cluster. As shown in Fig.~\ref{fig:lofar_filtered}, there is significant filamentary structure within the radio lobes at the connection regions with the jets and the southern bridge.

\begin{figure}
    \centering
    \includegraphics[width=\columnwidth]{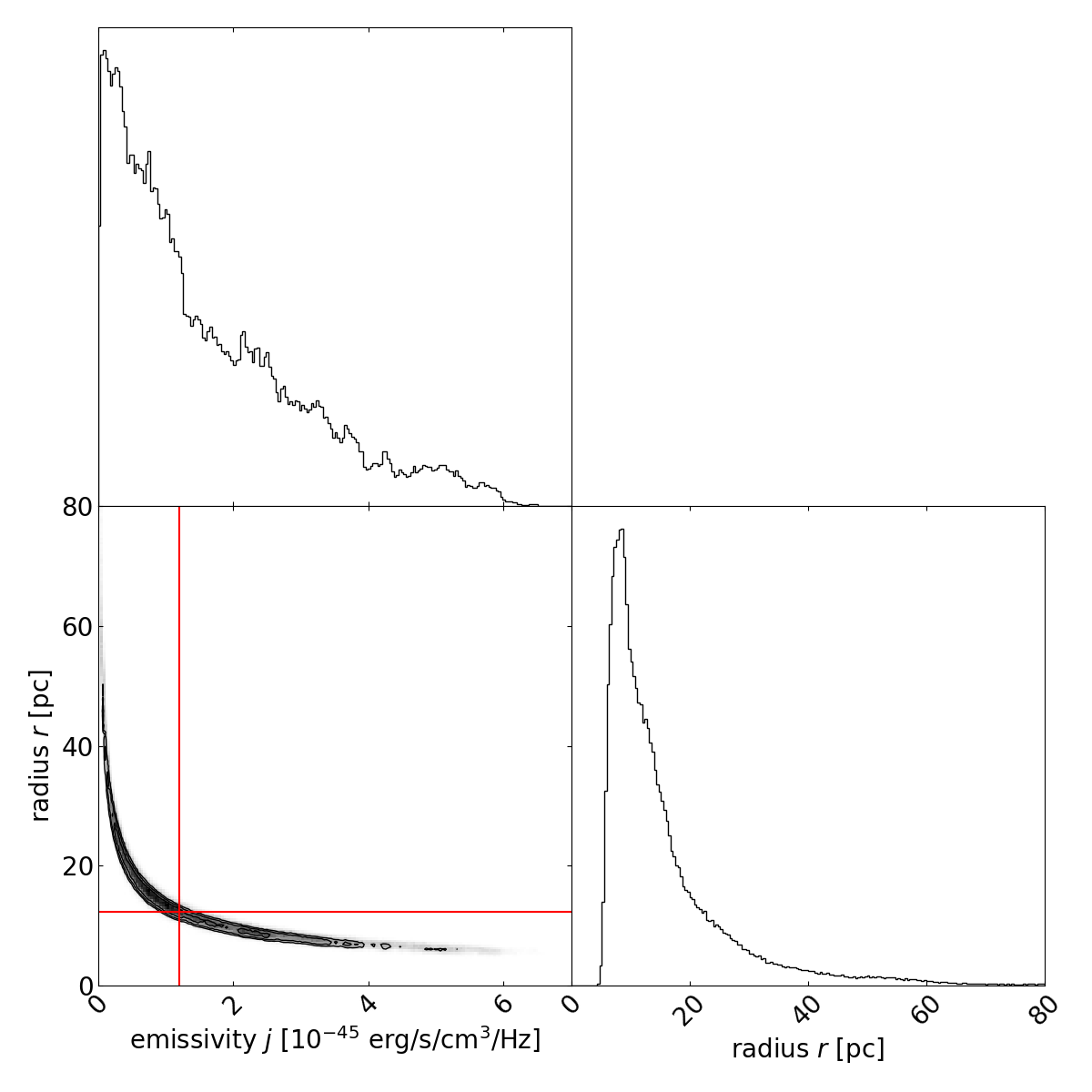}
    \caption{Corner plot showing the probability distributions for the radius of the isolated synchrotron threads and their emissivity, as obtained using MCMC fitting. The median value is indicated by the red crosshairs in the bottom-left panel.}
    \label{fig:mcmc_corner}
\end{figure}

Of key interest are the physical scale and emissivity of these synchrotron threads. Unfortunately, the angular scale of the threads is near or below the angular resolution of even our high-resolution LOFAR observations (\(\sim\)\(0.3^{\prime\prime}\rightarrow190\ \mathrm{pc}\)). To best extract the approximate physical properties of these synchrotron threads, we used a Markov-Chain Monte Carlo (MCMC) approach to fit a cylindrical shape convolved with the clean beam to the western bundle of threads. We selected two segments within the western threads that are relatively isolated from the remaining threads in order to obtain the most reliable results possible. These two segments are used as a proxy for the properties of the entire structure. Even with careful modelling, substantial systematic uncertainties should be taken into account when interpreting the results. From the MCMC, we obtain the distribution of emissivities and radii for these two isolated segments shown in Fig.~\ref{fig:mcmc_corner}. This shows that the 68.3\% confidence interval of the radius of the synchrotron threads is the range of \(8-24\) pc, with an emissivity in the range of \(0.3-3.2\times10^{-45}\ \mathrm{erg}/\mathrm{s}/\mathrm{cm}^{3}/\mathrm{Hz}\) at 144~MHz. We consider the lower limit on the scale of the threads, and likewise the upper limit on the emissivity, to not be meaningful as we can not reliably assume that the threads do not further break down into substructure. Therefore, we consider the primary result of this modelling to be that the physical radius of these threaded structures is most likely smaller than 24 parsecs and that the emissivity has to be at least \(0.3\times10^{-45}\ \mathrm{erg}/\mathrm{s}/\mathrm{cm}^{3}/\mathrm{Hz}\) in order to be consistent with the observations. Further details about the MCMC fitting are provided in Appendix~\ref{app:mcmc}.

\subsection{Spectral indices}

\begin{figure}
    \centering
    \includegraphics[width=\columnwidth]{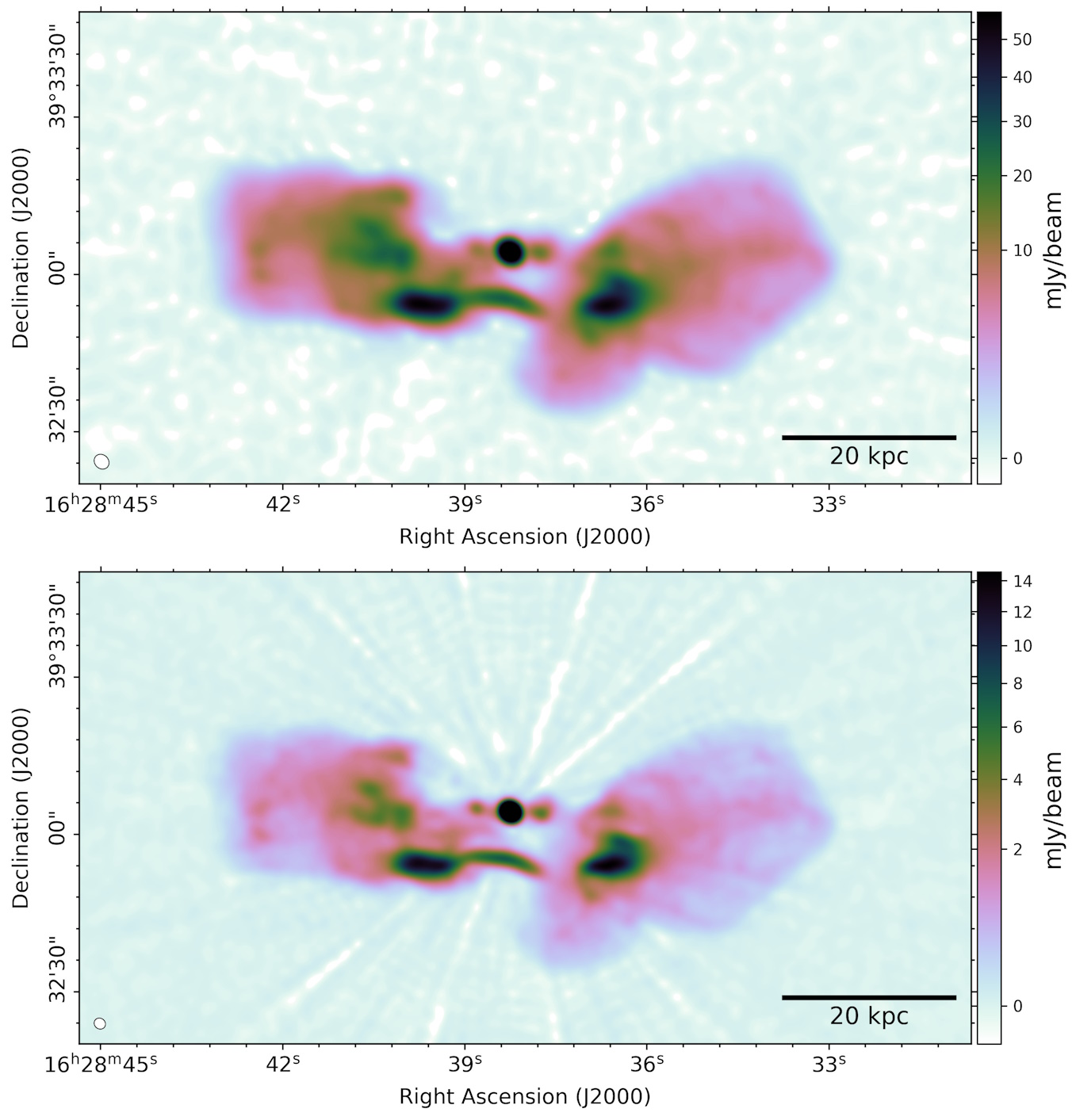}
    \caption{3C\,338 as imaged using the VLA at an observed frequency of 1.5~GHz (top panel) and 3~GHz (bottom panel). The scale bar in the bottom-right corner of each panel measures 20~kpc at the redshift of 3C\,338. The angular resolutions (1.5~GHz: \(3.030^{\prime\prime}\times2.619^{\prime\prime}, \mathrm{PA}=45.5^\circ\); 3~GHz: \(2.239^{\prime\prime}\times2.018^{\prime\prime}, \mathrm{PA}=53.3^\circ\)) are indicated by the ellipse in the bottom-left corner of both panels.}
    \label{fig:vla_maps}
\end{figure}

To provide a more complete view of 3C\,338, we also imaged the VLA L-band and S-band data, which show the radio emission at a frequency of 1.5~GHz and 3~GHz, respectively. Both images are shown in Fig.~\ref{fig:vla_maps}. Interestingly, the VLA images do not reveal any of the isolated synchrotron threads discovered by LOFAR. For the western threads, which are the brightest at 144~MHz, this non-detection in the 1.5~GHz VLA map implies an integrated spectral index of \(\alpha_{1500}^{144}<-3.0\) for the complete structure and a local spectral index of \(\alpha_{1500}^{144}<-3.6\) in the brightest part of the threads, based on the \(1\sigma\) rms noise level in the L-band image.

\begin{figure}
    \centering
    \includegraphics[width=\columnwidth]{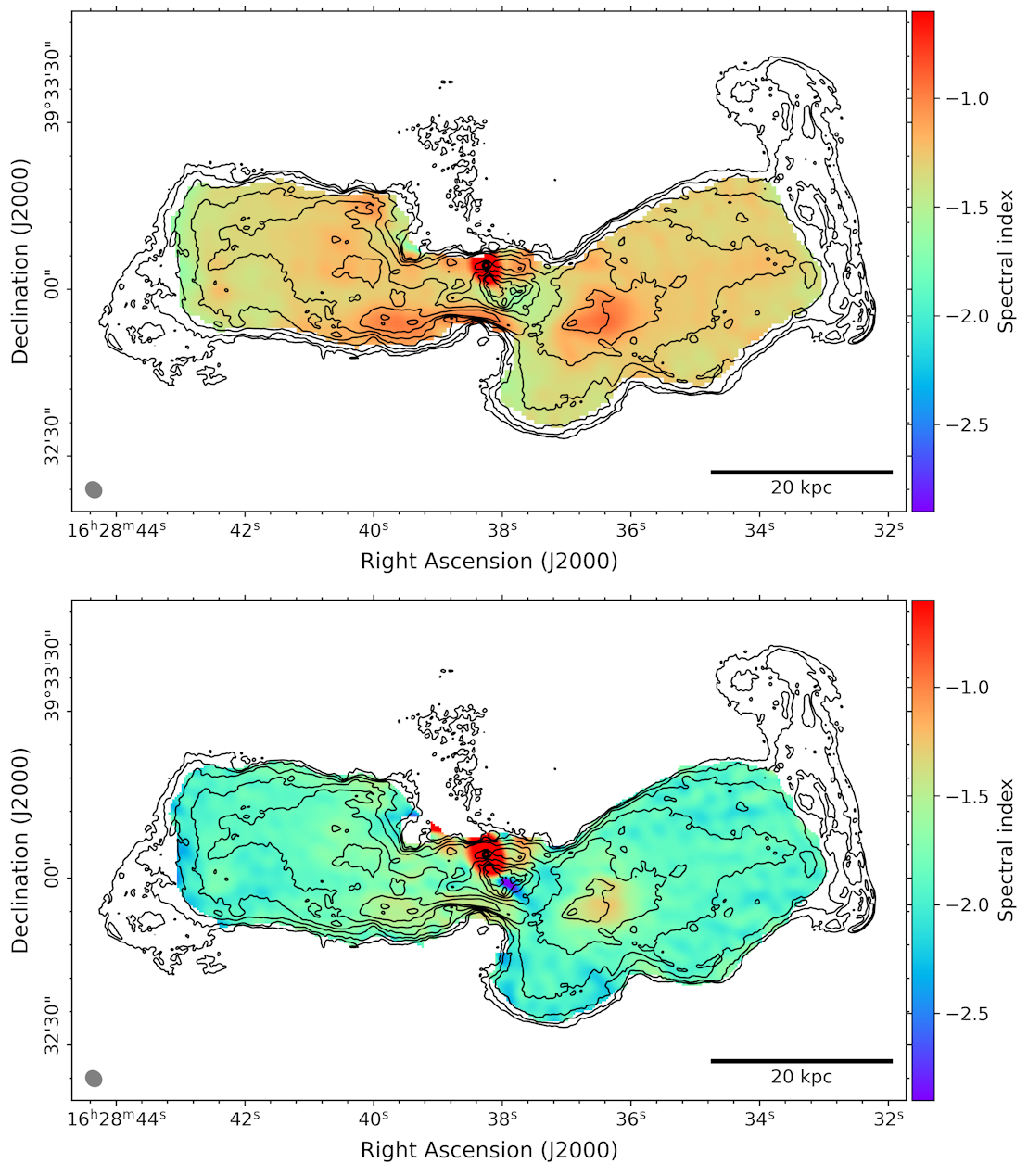}
    \caption{Spectral index maps of 3C\,338 between 144~MHz and 1.5~GHz (top panel) and 1.5~GHz and 3~GHz (bottom panel). The scale bar in the bottom-right corner of each panel measures 20~kpc at the redshift of 3C\,338. The grey ellipse in the bottom-left corner of each panel shows the angular resolution of the spectral index maps, respectively.}
    \label{fig:spixmaps}
\end{figure}

The spectral index of the radio lobes and the southern bridge can be mapped by combining the LOFAR 144~MHz map with the VLA L- and S-band maps, as shown in Fig.~\ref{fig:spixmaps}. These spectral index maps show a relatively spatially uniform spectral index in the radio lobes of approximately \(\alpha_{1500}^{144}\approx-1.3\pm0.1\) and \(\alpha_{3000}^{1500}\approx-1.9\pm0.1\). Interestingly, the southern bridge shows a flatter spectral index than the surrounding radio lobes, at around \(\alpha_{1500}^{144}\approx-1.0\pm0.1\) and \(\alpha_{3000}^{1500}\approx-1.5\pm0.1\).

\section{Discussion}

The high-resolution imaging of 3C\,338 at low radio frequencies has revealed multiple unusual features of this radio source. Isolated synchrotron-emitting threads are a recent feature of radio lobes to be discovered, with only a few known occurrences thus far. In this project, we targeted 3C\,338 due to the previously known radio bridge connecting the two radio lobes, and obtained serendipitous detections of prominent isolated synchrotron threads in multiple locations around and within the radio lobes. Additionally, the spectral index maps produced using LOFAR and VLA observations show almost no variation across the lobes. Here, we discuss the interpretation of these results.

\subsection{Nature of synchrotron threads}

The newly-discovered isolated synchrotron threads show a feature of radio lobes that has not been commonly observed before. Therefore, it is of interest to investigate how these structures came about and what kind of information they can provide about the interaction between active galactic nuclei and their environment. While we do not exclude that the southern bridge shares the same physical origin as the newly-discovered synchrotron threads, they exhibit substantial differences in physical scales and spectral indices. For that reason, we here discuss the newly-discovered isolated synchrotron threads separately to the southern bridge (see Sect.~\ref{sect:bridge}).

\subsubsection{Relation to the radio lobe}

The proximity between the isolated synchrotron threads and the two radio lobes poses the question whether the isolated synchrotron threads should be considered an extension of the radio lobes or a distinct structure. As previously mentioned and shown in Fig.~\ref{fig:lofar_filtered}, similar fine structure as in the isolated synchrotron threads can also be recognized within the radio lobes. It could be argued that the threads are either a direct extension of the radio lobes or simply show lingering radio emission from a past epoch of AGN activity. Such previous outbursts tend to be found farther away from the AGN and feature steep synchrotron spectra due to ageing \citep[e.g.,][]{timmerman22,ubertosi24}. This scenario can be tested using X-ray observations, which show the density of the intracluster medium (ICM) in the cluster core \citep{sarazin86}. As the radio lobes expand away from the AGN, they vacate the region from its ICM content, resulting in apparent ``bubbles'' or ``cavities'' within the ICM in X-ray images \citep{bohringer93,carilli94}. It is commonly found that previous outbursts from the AGN can be detected in the ICM even after their radio lobes fade away at higher frequencies due to synchrotron ageing, resulting in ``ghost cavities'' \citep[e.g.,][]{fabian06}. The radio emission within these cavities is then typically only found at sub-GHz radio frequencies, such as LOFAR's observing frequency \citep[e.g.,][]{vanweeren24}.

\begin{figure*}
    \centering
    \includegraphics[width=\textwidth]{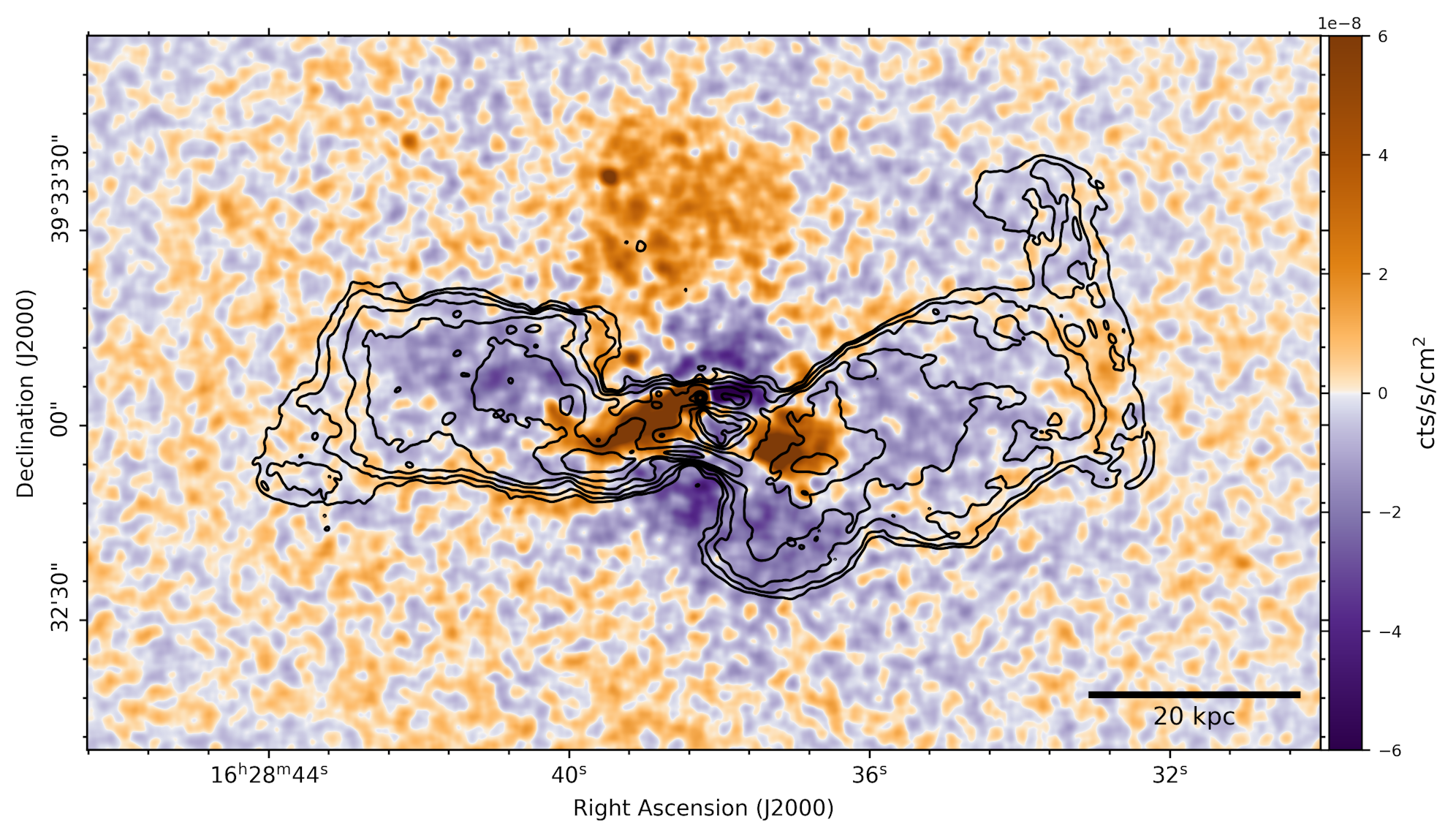}
    \caption{X-ray residual map of Abell\,2199 obtained by subtracting a smooth model for the ICM from the Chandra X-ray image. The contours indicate the radio emission at 144~MHz as observed by LOFAR. The scale bar in the bottom-right corner of each panel measures the listed physical length at the redshift of Abell\,2199}
    \label{fig:xray_overlay}
\end{figure*}

To test whether the isolated synchrotron threads can likely be considered part of a radio lobe, we correlated our LOFAR image with archival Chandra X-ray data. By subtracting a smooth ICM profile defined as a set of concentric Gaussians from the Chandra X-ray image, we obtain a residual map in which surface brightness excesses and depressions are emphasized. This residual map is shown in Fig.~\ref{fig:xray_overlay}, where ICM cavities can be identified as negative residuals. This residual map reveals a few key structures. Most importantly, the radio lobes directly match the X-ray cavities. The central core region shows some chaotic structure, likely caused by sloshing in the ICM \citep{nulsen13}, but beyond the central \(\sim\)10 kpc, the residuals show clear X-ray cavities that are coincident with the radio lobes. Critically, both the eastern and western synchrotron threads lie outside of the clear X-ray cavities. In the western synchrotron threads, a small X-ray excess can even be detected. A deficit in X-ray surface brightness can be found north of the western threads. If this is a cavity produced by a radio lobe emitted during a previous epoch of activity of the AGN, the relic cosmic ray electron population in this volume could help explain the hammerhead-like structure. We note that the eastern threads are small in scale, and therefore could realistically be associated with a cavity in the ICM that is too subtle to detect with the present observations. Notably, the northern thread appears to be directed towards a patch of excess X-ray surface brightness, which was found by \citet{nulsen13} to be a cooler region by about 0.5~keV. They suggest this northern plume is caused by ICM sloshing in the core. The association of this thread with the northern plume, as well as its location, strongly contradicts a radio-lobe origin of this thread. Based on their geometry, the correlation between the radio and X-ray emission, as well as the strong difference in spectral index between the two main radio lobes and the surrounding threads, we consider it most likely that the isolated synchrotron threads are separate structures from the two main radio lobes. Furthermore, because the threads are at least in part inconsistent with cavity-filling radio lobes, we suggest that the threads observed in 3C\,338 should be classified as a separate type of structure than a radio lobe, even though their cosmic ray electron population likely originates from radio lobes.

\subsubsection{Physical properties}

The isolated synchrotron threads detected in 3C\,338 show a number of differences and similarities to previously reported isolated threads, such as those in ESO\,137-006 \citep{ramatsoku20} and 3C\,40B \citep{rudnick22}. First of all, while all examples show compact and elongated structures, their morphologies are dissimilar. The case of ESO\,137-006 shows synchrotron threads connected to different radio lobes on either end. Meanwhile, the synchrotron threads in 3C\,40B are only connected to a radio lobe on one end and are then seen extending directly away from the radio lobe. In contrast, the western threads in 3C\,338 apparently run parallel to the surface of the radio lobe. Some threads appear to connect to the radio lobe, while others do not. The northern thread near the AGN is also unique in the sharp bend it experiences, which is not seen in such a way in most other threads. The eastern thread appears to loop around and connect back to the same radio lobe.

Furthermore, the western threads are unique in that they form a bundle. This bundle of threads consists of noteworthy features, such as the ``hammerhead''-like morphology at the northern end of the western threads. Here, it appears that the western threads connect perpendicular to a separate set of threads. Notably, within the main threaded bundle, the density and overall scale of the bundle varies. Additionally, the bundle cannot be resolved into individual threads everywhere. Instead, diffuse regions form, where threaded structures appear to merge together. Most clearly, directly south of the ``hammerhead'', there is a region where the density peaks and individual threads can no longer be distinguished. This suggests that the density of emission is primarily determined by the spacing between the threads. Alternatively, the threads could also feature intrinsic density variations along their length. Whereas the threads in ESO\,137-006 and 3C\,40B are measured to have a diameter of approximately 1~kpc, the threads in 3C\,338 show structure with a diameter of approximately \(\sim\)50~pc or less.

Finally, the spectral indices reported for these isolated synchrotron threads are also different from the examples in ESO\,137-006 and 3C\,40B. The threads in ESO\,137-006 and 3C\,40B feature relatively flat spectral indices, with a spectral index of \(\alpha_{1400}^{1000}\approx-2\) being reported for ESO\,137-006 \citep{ramatsoku20} and the spectral index for the threads in 3C\,40B varying between approximately \(-2.0<\alpha_{1670}^{900}<-1.0\) and \(-1.3<\alpha_{1283}^{144}<-0.8\) \citep{rudnick22}. In 3C\,338, we measure an upper limit on the integrated spectral index for the western threads of \(\alpha_{1500}^{144}<-3.0\) and a local spectral index within the brightest part of the threads of \(\alpha_{1500}^{144}<-3.6\). These constraints highlight the value of low-frequency radio observatories in searches for more examples of isolated synchrotron threads.

\subsubsection{Physical origin}

Threaded or filamentary emission has been previously observed in many different environments. Prominent examples include radio relics like Abell\,2256 \citep{rajpurohit22}, evolved AGN outbursts like Nest200047 \citep{brienza25}, and tailed radio galaxies like NGC\,1265 \citep{gendronmarsolais20} and in Abell\,2255 \citep{derubeis25}. Filamentary structures have even been reported in supernova remnants such as Vela X \citep{milne95} and Simeis\,147 \citep{khabibullin24}. However, isolated synchrotron threads such as the ones observed in ESO\,137-006 \citep{ramatsoku20} and 3C\,40B \citep{rudnick22} are a newly discovered phenomenon. Thanks to high-resolution low-frequency radio observatories, these displays of interaction between environmental magnetic fields and radio plasma now form an opportunity to probe the physical conditions within these structures. \citet{rudnick22} proposed that shear motions in the ICM could create magnetic threads. If brought in contact with the relativistic plasma from a radio jet or lobe, these magnetic threads embedded in the ICM could channel away and contain this plasma, resulting in isolated synchrotron threads. Given the strong sloshing observed in the ICM of Abell\,2199 \citep{nulsen13}, we find that this model could also feasibly apply to 3C\,338.

The prominent western threads in 3C\,338 are in particular interesting due to their morphology and spectral index. As previously discussed, these isolated threads are more compact than those previous reported, they feature a significantly steeper spectral index and they lie mostly tangential to the surface of the radio lobe rather than clearly extending away from the lobe. A substantial fraction of the isolated synchrotron threads also show no direct connection to the radio lobe which presumably initially fed them. We speculate that a likely explanation for these threads may be that they are significantly older threads than those in ESO\,137-006 and 3C\,40B, and initially formed connected to the radio lobe and extending away from the AGN. As the radio lobe then expands over time, it could compress the medium in which the synchrotron threads reside and push it to the sides of the radio lobe. This would cause the synchrotron thread to orient itself parallel with the surface of the radio lobe. If the isolated threads in 3C\,338 are relatively old examples, this would also be consistent with their spectral index, as synchrotron ageing typically results in steeper spectra.

\subsection{The southern bridge}\label{sect:bridge}

\begin{figure}
    \centering
    \includegraphics[width=\columnwidth]{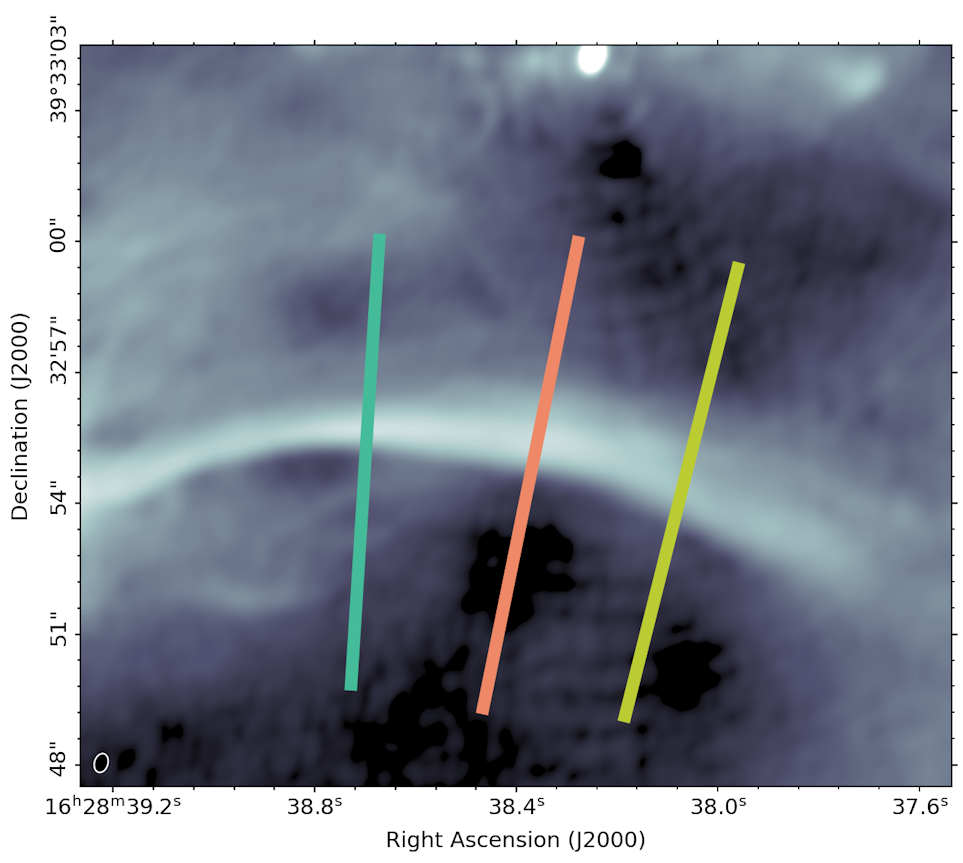}
    \includegraphics[width=\columnwidth]{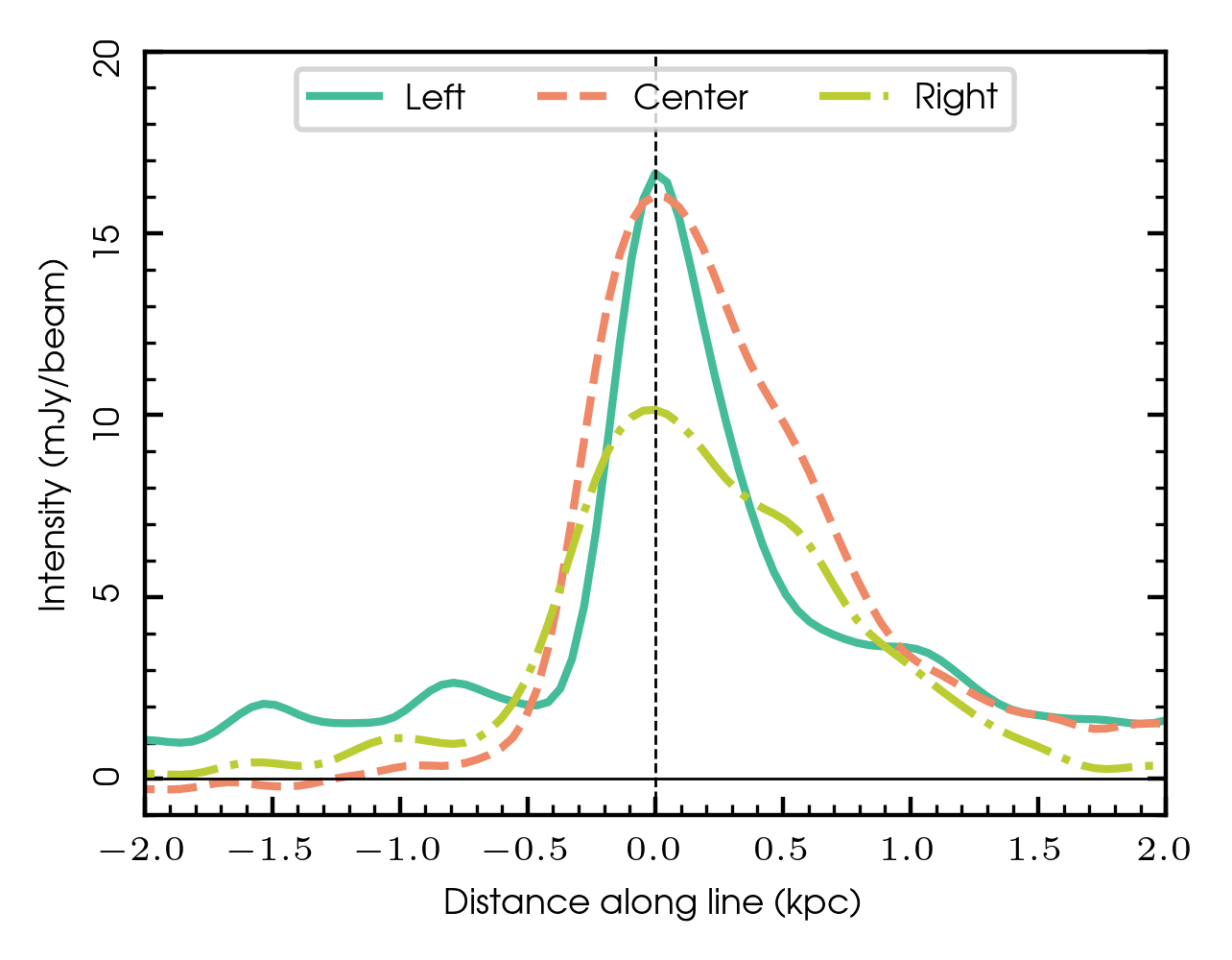}
    \caption{Three brightness profiles along the southern bridge in 3C\,338 extracted from the high-resolution International LOFAR Telescope image taken at 144~MHz. The top panel shows a zoomed-in view of the image in Fig.~\ref{fig:lofar_map} with the position of the three profiles indicated by the solid colored lines. The bottom panel shows the brightness measured along the profiles, with the solid teal line showing the left-most profile, the dashed orange line showing the center profile and the dash-dotted lime line showing the right-most profile. The distance measured along each profile was measured from south to north with respect to their brightness peaks.}
    \label{fig:profiles}
\end{figure}

The nature of the southern bridge has been a topic of discussion since its discovery by \citet{burns1983}. This bridge resembles a set of radio jets in practically every way except in that it lacks an AGN along its path. \citet{burns1983} proposed two explanations: (i) that the bridge is created as the radio lobes experience ram pressure from the cooling ICM from the south, or (ii) that the bridge is an aged jet originally produced by the AGN detected directly north of it. However, both of these models fail to provide a compelling explanation.

As indicated by \citet{burns1983}, the ICM-cooling model would only work with a highly asymmetric cooling flow, which is unlike what is normally observed in galaxy clusters. Additionally, the steep spectrum at high frequencies suggests that the radio plasma in the bridge has not experienced any (re-)acceleration, although this would be expected from the collision with such a cooling flow. Remarkably, one prediction of the ICM-cooling model is consistent with the new LOFAR observations. The ICM-cooling model predicts an asymmetric brightness profile across the bridge, with the brightness peaking towards the south. As shown in Fig.~\ref{fig:profiles}, this is in fact detected in the high-resolution LOFAR map.

In the aged-jet model, where they assumed that the AGN's host galaxy migrated from the bridge to its current position, there are also a number of observational challenges. Primarily, such a migration due to orbiting motion would realistically require \(\sim\)\(10^7\) years, and it is unlikely that the plasma in the old jet would not have yet travelled to the radio lobes, which are only \(\sim\)\(10\ \mathrm{kpc}\) away.

\citet{gopalkrishna24} considered an entirely different physical origin for synchrotron threads connecting two radio lobes. They present a model in which the synchrotron treads, such as those observed in between two radio lobes, are caused by electric discharges along the jets. 

Critically, we note that none of the previously mentioned models would be able to explain at least the eastern and western threads here reported in 3C\,338. The newly-discovered isolated threads extend away from the rest of the rest of the structure, which cannot be explained by the inwards-pointing pressure created by an ICM cooling flow. Similarly, both the aged-jet model and the electric-discharge models cannot explain the eastern and western threads, which lie on the opposite side of the radio lobes and do not extend between the radio lobes. Therefore, these models for the southern bridge in 3C\,338 would imply that the southern bridge has a different physical origin than least the eastern and western threads.

A model that would unite the southern bridge with the newly-discovered isolated synchrotron threads is the magnetic thread model proposed by \citet{rudnick22} for 3C\,40B. The shear motions required to form such magnetic threads are likely to be sufficiently present due to the intense sloshing observed in the ICM \citep{nulsen13}. Assuming the radio plasma within the magnetic threads ages due to synchrotron emission, this would also be consistent with the spectral indices observed within 3C\,338. 
However, one issue is that the spectral index observed in the southern bridge is flatter than that of the radio lobes on either side, which is opposite to what is observed between the radio lobes and the western threads. Therefore, it is difficult to unite the southern bridge with the magnetic threads within this model.

\subsection{Anomalous spectral indices of 3C\,338}

Some of the spectral index patterns in 3C\,338 are unexpected given standard spectral ageing models. At first glance, the fact that the southern bridge, which has no obvious source of fresh electrons, is flatter than the lobes may appear unexpected. However, this is likely simply due to the curved electron spectrum radiating in a stronger magnetic field. The main unexpected element, however, is that the diffuse radio lobes show very little variation from one end to the other, unlike what would be expected if the far end of the radio lobes contained older radio plasma.

\subsubsection{Spectral structure}\label{sect:spectralstructure}

\begin{figure}
    \centering
    \includegraphics[width=\columnwidth]{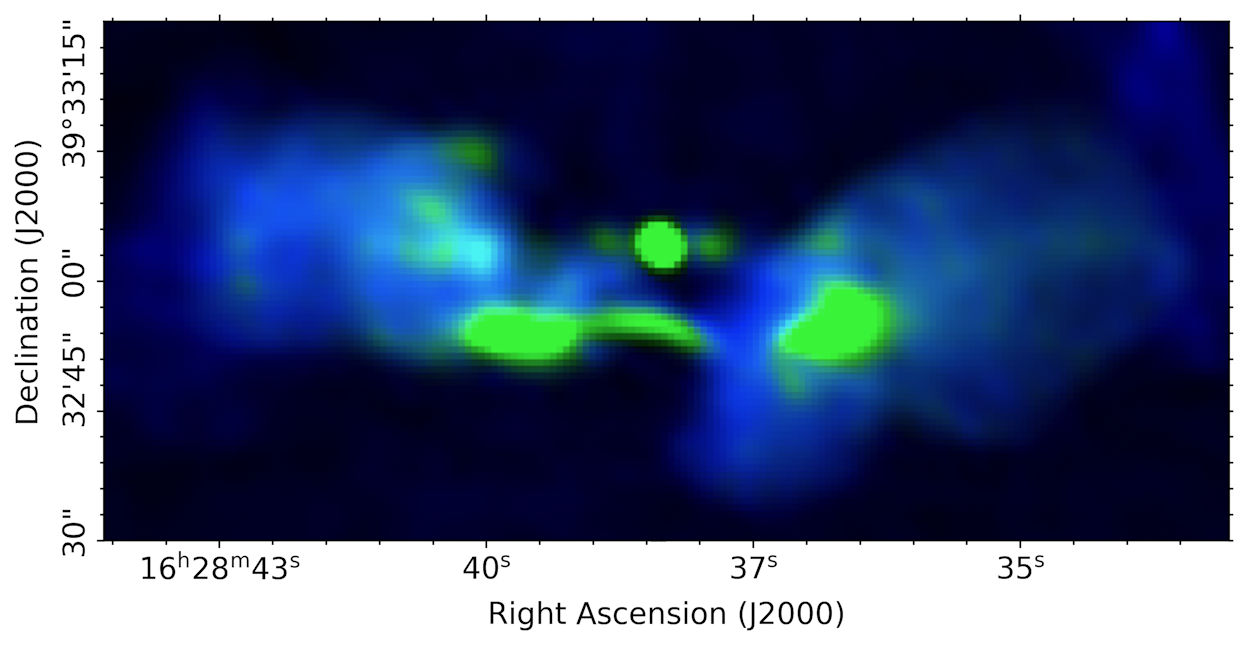}
    \caption{Spectral tomography image of 3C\,338. The 1500~MHz image is shown in green, scaled to emphasize the bright features. In blue is the tomography image corresponding to \(S_{144~\mathrm{MHz}} - 13.5 \times S_{1500~\mathrm{MHz}}\), where \(S_\nu\) is the surface brightness at an observed frequency \(\nu\). Any structures in the blue image, whether overlapping with others or not, will be bright (or disappear, or be over-subtracted), for $\alpha < (\mathrm{or} =, \mathrm{or} >)~-1.1$ .}
    \label{fig:tomography}
\end{figure}

A more detailed visualization of the spectral index structure can be seen through ``spectral tomography'' \citep{rudnick96}, which preserves (some of) the total intensity information, while highlighting spectral variations. Figure~\ref{fig:tomography} shows that the flatter regions are those with brighter, small scale structures, in comparison with the diffuse, steeper emission. As seen in Fig.~\ref{fig:spixmaps}, there is no monotonic trend for steeper regions to be further from the host, as might be expected if there were increased losses with distance from the host. 

\begin{figure}
    \centering
    \includegraphics[width=\columnwidth]{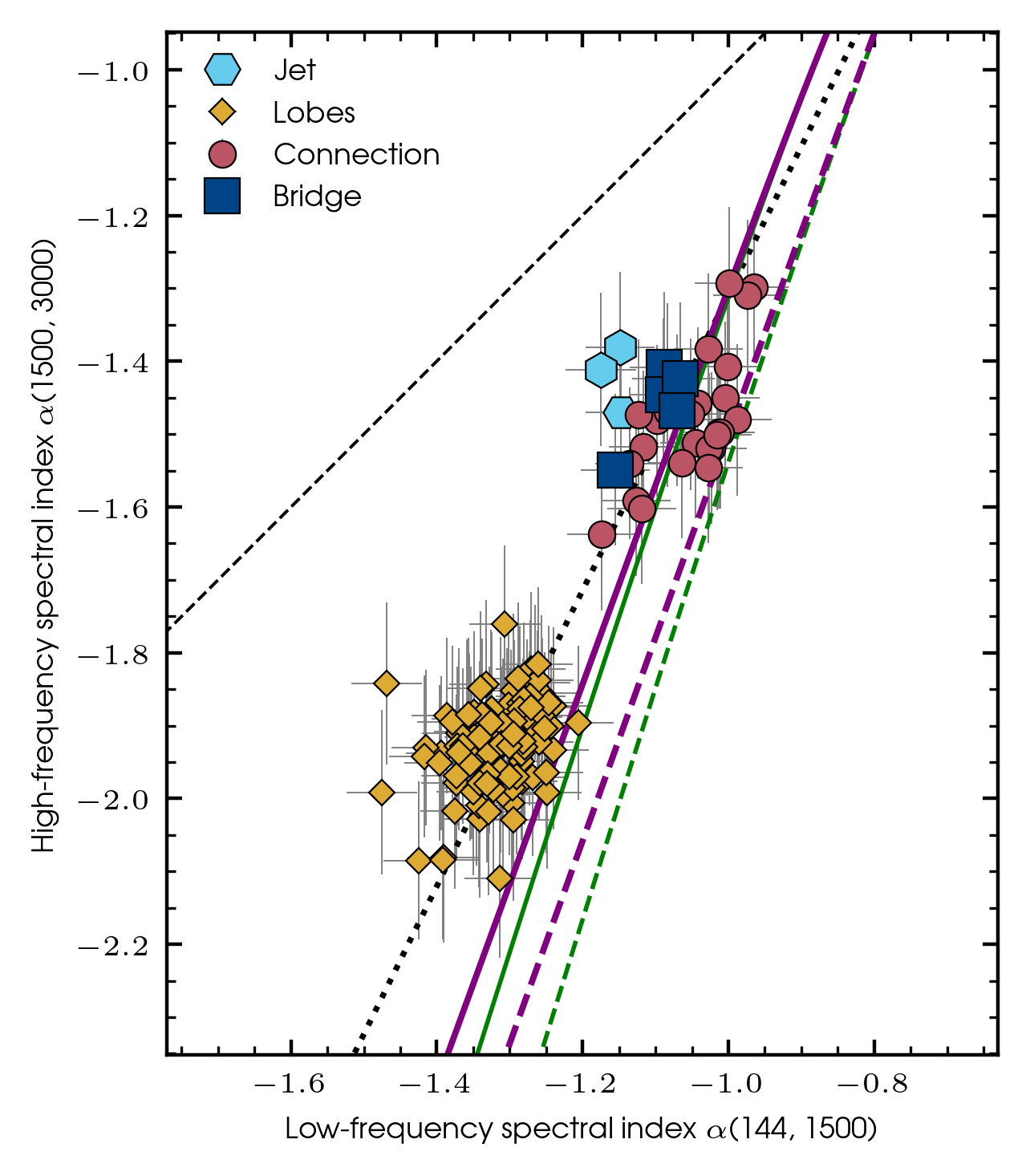}
    \caption{Color-color diagram for the radio emission in the jets of 3C\,338, the radio lobes, the southern bridge, and the region within the radio lobe where the southern bridge connects. The light blue hexagons sample the jetted regions immediately east and west of the AGN core. The dark blue squares sample a selection of points within the bridge. The red circles sample the region within the radio lobe where the bridge connects, and the yellow diamonds sample the remaining regions of the radio lobes. The green lines shows the spectral ageing of a Jaffe-Perola model and the purple lines shows the spectral ageing following a Tribble JP model. The dashed versions of these lines show these models with an injection index of \(\alpha_\mathrm{inj}=-0.8\), while the solid versions of the green and purple lines show these model with an injection index of \(\alpha_\mathrm{inj}=-0.7\). The black dotted line shows the color-color relation used for the emissivity calculations.}
    \label{fig:colorcolor}
\end{figure}

With three frequencies, we can explore the shape of the underlying spectra using a color-color diagram, which plots the high-frequency spectral index of individual beams or regions vs. the corresponding low-frequency spectral index. If the underlying electron populations have a unique shape throughout the source, and differ only in their local magnetic field strengths or in their cut-off energies (as would be due to various amounts of radiative or adiabatic losses), then the points will follow a single locus of points in color-color space. Figure~\ref{fig:colorcolor} shows the spectra of regions of the radio lobes and the southern bridge. We find that all of the radio emission is consistent with such a single locus. The regions within the southern bridge and the connection site lie at the flat-spectrum end of the distribution, while the radio lobe regions are steeper.

Various lines in the color-color diagram show the behaviors expected from different spectral (electron) shapes. Increasing steepness corresponds to lower cut-off frequencies. We find that the data are not consistent with the standard Jaffe-Perola \citep[JP; ][]{jaffe73} or Tribble JP \citep{tribble93} spectra, but are well-fit by the empirical form 
\begin{equation}
    S(\nu_0) = \left(\nu_0/\nu_c\right)^{\alpha_0}e^{-\sqrt{\nu_0/\nu_c}},
\end{equation}
with \(\alpha_0=-0.7\). This distribution is smoother (lower curvature) than the JP spectrum, and does not correspond to any of the standard idealized theoretical forms. However, it can easily be reproduced by mixing different JP spectra, for example, over a range of \(\sim\)\(50\%\) in \(\nu_c\).  This could result from variations in magnetic field or in losses within the sampled regions. 

We can further explore the causes of the spectral variations by looking at their dependence on the observed emissivities. This is similar to the analysis of \citet{katz-stone94}, which attempts to isolate which of the underlying physical parameters are responsible for the observed variations in brightness and spectral index across the source. The emissivity of a feature at a fixed observing frequency will depend on the number density of relativistic electrons (\(n\)), the local magnetic field (\(B\)), and the local electron energy cut-off, (\(\gamma_c\)). The observed spectral index depends only on the last two quantities. This type of analysis assumes that the different regions of the source have the same-shaped distribution of relativistic electrons; that shape is not affected by radiative losses or changes in magnetic field strength or adiabatic changes. As we have shown in Fig.~\ref{fig:colorcolor}, this is a good assumption for 3C\,338. 

If variations in the electron number density were the dominating factor between regions, then there would be no relationship between the local emissivity (brightness) and spectral index. Changes in either of the other two parameters would force a correlation between emissivity and spectral index. Since there is a correlation, as seen in Fig.~\ref{fig:tomography}, we start with the simplifying assumption that the number density is constant, and ask how variations in the other two factors would appear. Then, we can rewrite the above expression for spectral shape in terms of the emissivity (\(\epsilon\)) at frequency \(\nu\) for some region in the source as
\begin{equation}\label{eq:emm}
    \epsilon\left(\nu,\gamma_c,B\right)=C\cdot\left(\frac{\nu}{B\gamma_c^2}\right)^{\alpha_0}\mathrm{exp}\left(-\sqrt{\frac{\nu}{B\gamma_c^2}}\right).
\end{equation}

We can then vary \(B\) or \(\gamma_c\) and calculate the resulting emissivity.  Then, evaluating this expression at 144, 1500, and 3000~MHz allows us to calculate \(\alpha_{144}^{1500}\) and \(\alpha_{1500}^{3000}\) and thus the relationship between those spectral indices and the emissivity for various regions within the source.

\begin{figure}
    \centering
    \includegraphics[width=\columnwidth]{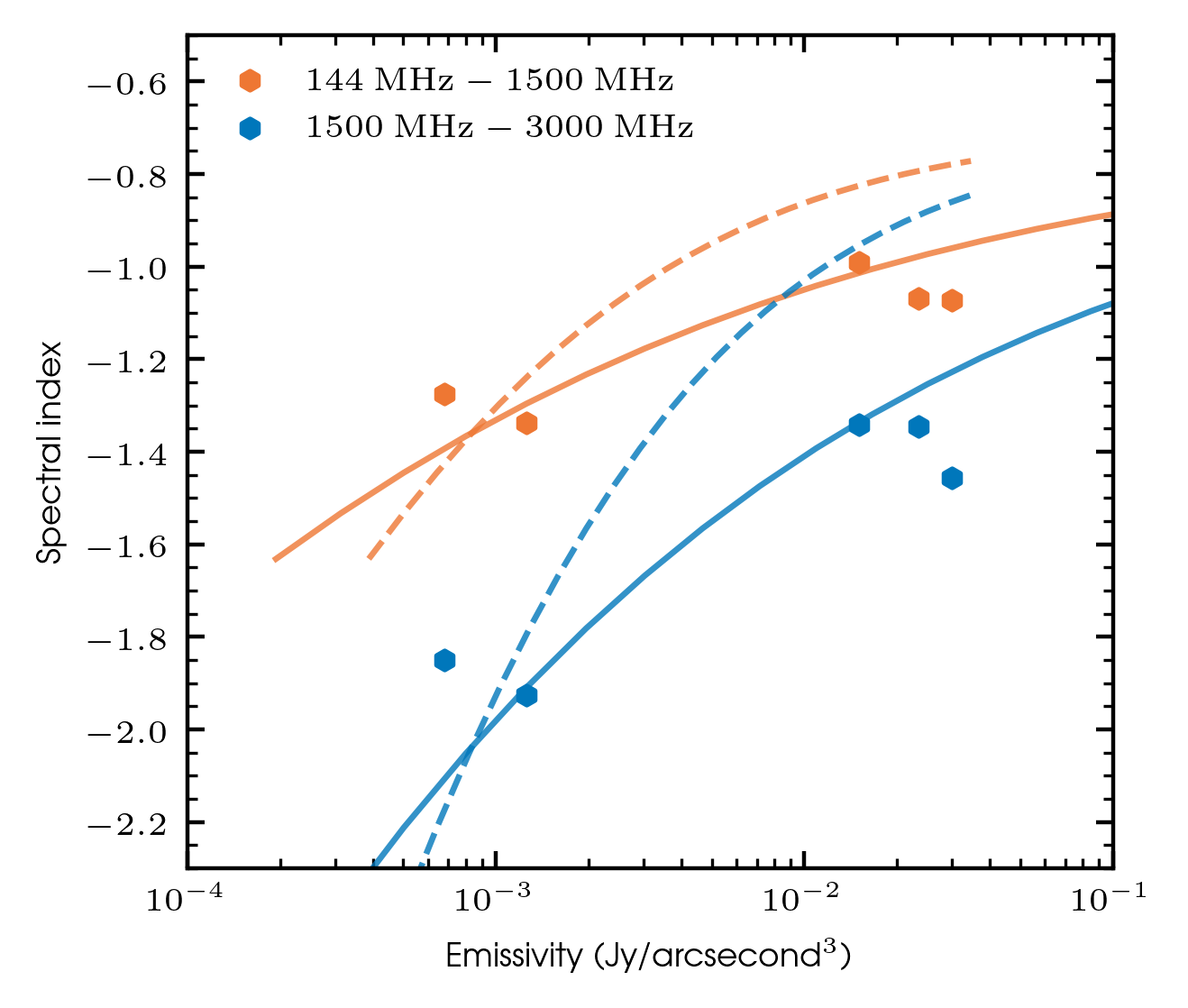}
    \caption{Spectral index as a function of emissivity at 144~MHz. The orange hexagons indicate the measurements for \(\alpha_{144}^{1500}\) and \(\alpha_{1500}^{3000}\), respectively. The dashed lines show the model varying only the cut-off value (\(\gamma_c\)) in Eq.~\ref{eq:emm}, such as due to radiative losses, while the solid lines show the model for varying only the magnetic field strength (\(B\)).}
    \label{fig:scaling}
\end{figure}

We took five representative regions within the source, and evaluated their emissivity at 144~MHz. These regions are columns taken through the center of the southern bridge, the eastern and western lobes, the connection hotspot in the western lobe, and the jet propagating away from the AGN. For these columns, we estimated the emissivity as the surface brightness divided by an estimate of the depth of the column. These are plotted in Fig.~\ref{fig:scaling} against the spectral indices measured in these regions, along with the expected relationships when either \(B\) or $\gamma_c$ are varied. While there are 10 independent observed pairs of values, each model has only one free parameter, either $\gamma_c$ or \(B\). The data appear reasonably close to the expectations from variations in \(B\) only, and inconsistent with variations in $\gamma_c$. This suggests that the spectral variations in the radio lobes and southern bridge of 3C\,338 are dependent primarily on differences in local magnetic field strength rather than by radiative losses.

\subsubsection{Spectral ageing estimates}\label{sect:spectralage}

We used the Broadband Radio Astronomy Tools \citep[BRATS;][]{harwood13,harwood15} package to estimate approximate spectral ages accounting for inverse Compton and synchrotron losses, and assuming equipartition magnetic field strengths, and a JP spectrum. More detailed models with additional free parameters to account for the exact spectral shape are not justified given the other simplifying assumptions used here.

Following \citet{govoni04} and \citet{beck05}, we obtain an equipartition magnetic field strength of 21.4~\(\mu\)G, assuming an injection index of \(\alpha_\mathrm{inj}=-0.7\), a value of \(K_p = U_\mathrm{protons}/U_\mathrm{electrons}=1\), and a low-energy cut-off on the cosmic ray particle energy distribution of \(\gamma_\mathrm{min}=100\). Using this magnetic field strength, and a spectral index in the radio lobe at the side nearest to (or farthest from) the AGN of \(\alpha^{1500}_{3000}\approx-1.8\) (or \(-2.0\)), we calculate a spectral age of 9.4 (or 10.4)~Myr. These estimates suggest that the radio plasma in the outer region is only 1.0~Myr older than that in the inner regions. Given the scale of the radio lobes of approximately 30~kpc, this would require the radio plasma to be transported at a velocity of at least 29,400~km/s, or around \(0.1c\).

A lower limit to the electron transport velocity can be obtained by assuming the maximum lifetime magnetic field, given by \(B_\mathrm{CMB}/\sqrt{3} = 2\)~\(\mu\)G \citep{stroe14} instead of its equipartition value. Assuming this magnetic field strength throughout the source and constant in time, this results in a difference in spectral age between the near-side and far-side of the radio lobes of 9.4~Myr, and a corresponding velocity of 3,100~km/s, or about \(0.01c\).

\subsubsection{Homogenizing the electrons in 3C\,338's lobes}

If the field within 3C\,338's lobes is the maximum lifetime field of 2~\(\mu\)G, then the required 3100 km/s velocities are quite reasonable. Such velocities would equate to a Mach number of around 3-4 \citep{johnstone02,nulsen13}, which is within the range observed for radio lobes in galaxy clusters \citep{ineson17}. However, assuming equipartition, our observations predict a substantially higher magnetic field strength, which is in line with typical magnetic field strengths observed in cool-core galaxy clusters \citep[e.g.,][]{carilli02}. We therefore explore alternative ways to homogenize the population if the field were approximately its equipartition value of \(\sim21\)~\(\mu\)G.

One possibility would be to eliminate the far-near difference in spectra by assuming that the lobes were formed through the spherical expansion of plasma from a compact region where the jet from the nucleus dumped all of its energy near the centers of the current lobes. This, however, would result in a dramatic steepening towards all the edges of the lobes because of adiabatic losses affecting $\gamma_c$ and \(B\), which is not observed. It would also not explain the observed correlation between the emissivity and the spectral index, which simply requires a change in \(B\), not $\gamma_c$ in addition. Coupled with the adiabatic reductions in \(n\), we would also expect very large decreases in emissivity towards the edges, which is not observed.

Alternatively, we can explore two mechanisms of homogenizing the relativistic plasma at speeds of at least \(0.1c\): a) bulk motions originating from the jet's momentum, or b) transport via Alfv{\'e}n waves. In addition, we also require that the medium within the radio lobes must provide sufficient pressure to support the lobes against the confining pressure of the ICM. In the cool core of Abell\,2199 at the radius of the lobes, the ICM density is \(n=0.0157\ \mathrm{cm}^{-3}\) at a temperature of \(kT=3.2\ \mathrm{keV}\) \citep{johnstone02,vacca12}, corresponding to a pressure of \(1.55\times10^{-10}\ \mathrm{erg/cm}^3\). At equipartition, the relativistic plasma itself provides only \(3\times10^{-11}\ \mathrm{erg/cm}^3\). We consider two ways to provide the needed pressure, either a quasi-static thermal component or one that flows through the lobe volume at considerable velocities providing kinetic pressure.

In the quasi-static case, transport mechanism a), bulk motion, is not relevant by definition. To estimate the plausibility of mechanism b) Alfv{\'e}n wave transport, we start by assuming the extreme situation where the thermal density in the lobes is that same as the external density. The Alfv{\'e}n speed is then \(v_A = B/\sqrt{\mu_0\rho}=4.6\times10^7\ \mathrm{cm/s}\), a factor of \(\sim\)60 less than the \(0.1c\) required to homogenize the electrons. If quasi-static pressure were provided instead by a much lower density  (e.g., \(2\times10^{-6}\ \mathrm{cm}^{-3}\)), but hotter medium, then this would increase the Alfv{\'e}n speed to the required levels. Modeling of the X-ray cavities associated with the lobes by \citet{vacca12} shows that they are consistent with being completely devoid of thermal gas. However, the low density cited above would require an unrealistically high temperature of \(10^6\)~keV to provide the necessary static pressure. The quasi-static case was similarly tested on MS\,0735.6+741 by \citet{abdulla19} using Sunyaev-Zel'dovich observations, which provided comparable results. Therefore, we find that a quasi-static thermal component within the lobes cannot plausibly homogenize the relativistic electrons.

Next, we consider whether the required pressure in the lobes could be provided by bulk motions such as a backflow or from other circulating motions within low-density lobes. Assuming a typical lobe advance speed of \(0.01c\) \citep[e.g.,][]{nulsen05,croston09,ineson17}, we can approximate the magnitude of the bulk motions within the radio lobe based on the ratio between the major and minor axes of the radio lobes. For 3C\,338, this ratio is approximately 2, implying bulk motions within the lobes on the order of 1500~km/s. In order to provide the pressure required to sustain the radio lobe against the external ICM pressure, a thermal density of \(6.6\times10^{-3}\ \mathrm{cm}^{-3}\) is required. This is only a factor of two below the external thermal density, which is ruled out by X-ray observations \citep{vacca12}, so mechanism a) bulk transport, is again ruled out. Similarly, the associated Alfv{\'e}n speeds of \(0.002c\) would also be insufficient to homogenize the lobes.

If we push the bulk motion possibility to an extreme, considering velocities within the lobe of  order \(0.1c\), then the required thermal density would only be around \(2\times10^{-5}\ \mathrm{cm}^{-3}\). This satisfies the X-ray observations and provides the required pressure in the radio lobes. Additionally, these densities would enable an Alfv{\'e}n speed of \(0.05c\), which is substantial enough for Alfv{\'e}n waves to contribute to the homogenization of the radio lobes. However, while these very fast bulk motions can provide both sufficient pressure and high transport speeds, they significantly exceed typical estimates \citep[e.g.,][]{stimpson23,dutta24}.

In summary, we find that we can rule out the scenarios of a rapid expansion of the radio lobes or a static thermal component. The most probable models suggest that the homogenous relativistic plasma in the radio lobes of 3C\,338 requires either a substantial deviation from equipartition (the \(0.01c\) case) or substantially higher bulk motions than normally observed within radio lobes in galaxy clusters (the \(0.1c\) case).

\section{Conclusions}

In this paper, we presented new high-resolution low-frequency radio maps of 3C\,338 made using the International LOFAR Telescope. These high-resolution LOFAR maps have revealed the presence of multiple narrow isolated synchrotron threads: (i) an eastern thread connecting to the eastern radio lobe on both ends, (ii) a northern thread connected to the inside of the western lobe and directed first towards the AGN and later directly north, and (iii) a bundle of threads located at the far end of the western lobe. Based on the non-detection of these threads in higher-frequency radio maps obtained using the VLA, we can constrain the integrated spectral index of the western threads to be \(\alpha_{1500}^{144}<-3.0\) and local spectral index within the brightest region of these threads to be \(\alpha_{1500}^{144}<-3.6\). Using MCMC fitting, we estimate that the physical radius of these threads is \(\lesssim24\) pc, though we emphasize that such measurements should be carefully interpreted as they are extracted near or below the angular resolution of the interferometer. The X-ray image taken with Chandra suggests that these synchrotron threads reside outside of the cavities in the ICM. Although the western threads could represent old radio lobes, we favor the model proposed by \citet{rudnick22}, where synchrotron threads follow magnetic threads created by sloshing in the ICM, to explain the newly discovered isolated threads in 3C\,338.

Furthermore, our high-resolution maps revealed complex filamentary structure within the radio lobes as well as a detailed view of the bridge of radio emission south of the AGN core. We considered whether the southern bridge can share the same physical origin as the newly-discovered synchrotron threads and find that none of the previously proposed models for this bridge can explain both the bridge and the threads. Alternatively, the southern bridge could be considered to be an example of a magnetic thread, but it is unclear how this can be unified with the spectral index of the bridge, which is flatter than that of the radio lobes.

Finally, our combination of LOFAR and VLA observations reveals that the spectral indices of the radio lobes are remarkably uniform, and that the jet, lobes and bridge are all consistent with a homogeneous electron population radiating in different local magnetic fields. The observed homogenization in the radio lobes could feasibly be achieved either by a substantial deviation from equipartition with standard lobe velocities or through high bulk motions in the lobes supported by a low-density thermal plasma. We rule out that the radio lobes are supported by a practically static thermal component with the homogenization achieved through Alfv{\'e}n waves or that the lobes experienced a period of rapid expansion.
    
%%%%%%%%%%%%%%%%%%%%%%%%%%%%%%%%%%%%%%%%%%%%%%%%%%

\section*{Acknowledgements}

We thank the anonymous referee for their helpful comments.
RT is grateful for support from the UKRI Future Leaders Fellowship (grant MR/Y020405/1). This work was supported by the STFC [grants ST/T000244/1, ST/V002406/1].
This paper is based (in part) on data obtained with the International LOFAR Telescope (ILT) under project code LC14\_019. LOFAR \citep{haarlem13} is the Low Frequency Array designed and constructed by ASTRON. It has observing, data processing, and data storage facilities in several countries, that are owned by various parties (each with their own funding sources), and that are collectively operated by the ILT foundation under a joint scientific policy. The ILT resources have benefitted from the following recent major funding sources: CNRS-INSU, Observatoire de Paris and Université d'Orléans, France; BMBF, MIWF-NRW, MPG, Germany; Science Foundation Ireland (SFI), Department of Business, Enterprise and Innovation (DBEI), Ireland; NWO, The Netherlands; The Science and Technology Facilities Council, UK; Ministry of Science and Higher Education, Poland. This work made use of the Dutch national e-infrastructure with the support of the SURF Cooperative using grant no. EINF-1287. This project has received support from SURF and EGI-ACE. EGI-ACE receives funding from the European Union’s Horizon 2020 research and innovation programme under grant agreement No. 101017567. The Jülich LOFAR Long Term Archive and the German LOFAR network are both coordinated and operated by the Jülich Supercomputing Centre (JSC), and computing resources on the supercomputer JUWELS at JSC were provided by the Gauss Centre for Supercomputing e.V. (grant CHTB00) through the John von Neumann Institute for Computing (NIC). 

%%%%%%%%%%%%%%%%%%%%%%%%%%%%%%%%%%%%%%%%%%%%%%%%%%

\section*{Data Availability}

All observations used in this article are publicly available through their respective archives. The LOFAR data can be found in the LOFAR Long Term Archive under project code LC14\_019. The VLA data can be accessed through the NRAO Data Archive under project codes 18B-159 and 19A-132 (see Table~\ref{tab:VLA}). The Chandra data can be obtained from the Chandra Data Archive using ObsIDs 497, 498, 10748, 10803, 10804 and 10805. All derived data products and scripts will be shared upon reasonable request to the lead author.


\bibliographystyle{mnras}
\bibliography{refs}

%%%%%%%%%%%%%%%%%%%%%%%%%%%%%%%%%%%%%%%%%%%%%%%%%%

%%%%%%%%%%%%%%%%% APPENDICES %%%%%%%%%%%%%%%%%%%%%

\appendix

\section{Markov-Chain Monte Carlo modelling}\label{app:mcmc}

In the interest of fully disclosing the results of our Markov Chain Monte Carlo (MCMC) simulations, we here present the details of the methodology as well as the best-fitting results. We assumed that the threads can be described as long cylinders of uniform emissivity, and fixed the position and shape of these cylinders following the threads visible in the image. We then created a model image by integrating the emissivity along the line of sight in each pixel, assuming that the synchrotron threads are optically thin. For improved accuracy, this model image was oversampled by a factor of 4 per axis, for a total of 16. We tested whether the results significantly changed by increasing the oversampling, but found that higher oversampling factors converged to the same probability distributions. This model was then convolved with the clean beam of the LOFAR-VLBI observation and averaged down to the pixel scale of the original image to produce an simulated thread. After subtracting this thread from the original image, we calculated a reduced \(\chi^2\) score based on the region within 150~pc of the center line of the thread.

\begin{figure}
    \centering
    \includegraphics[width=\columnwidth]{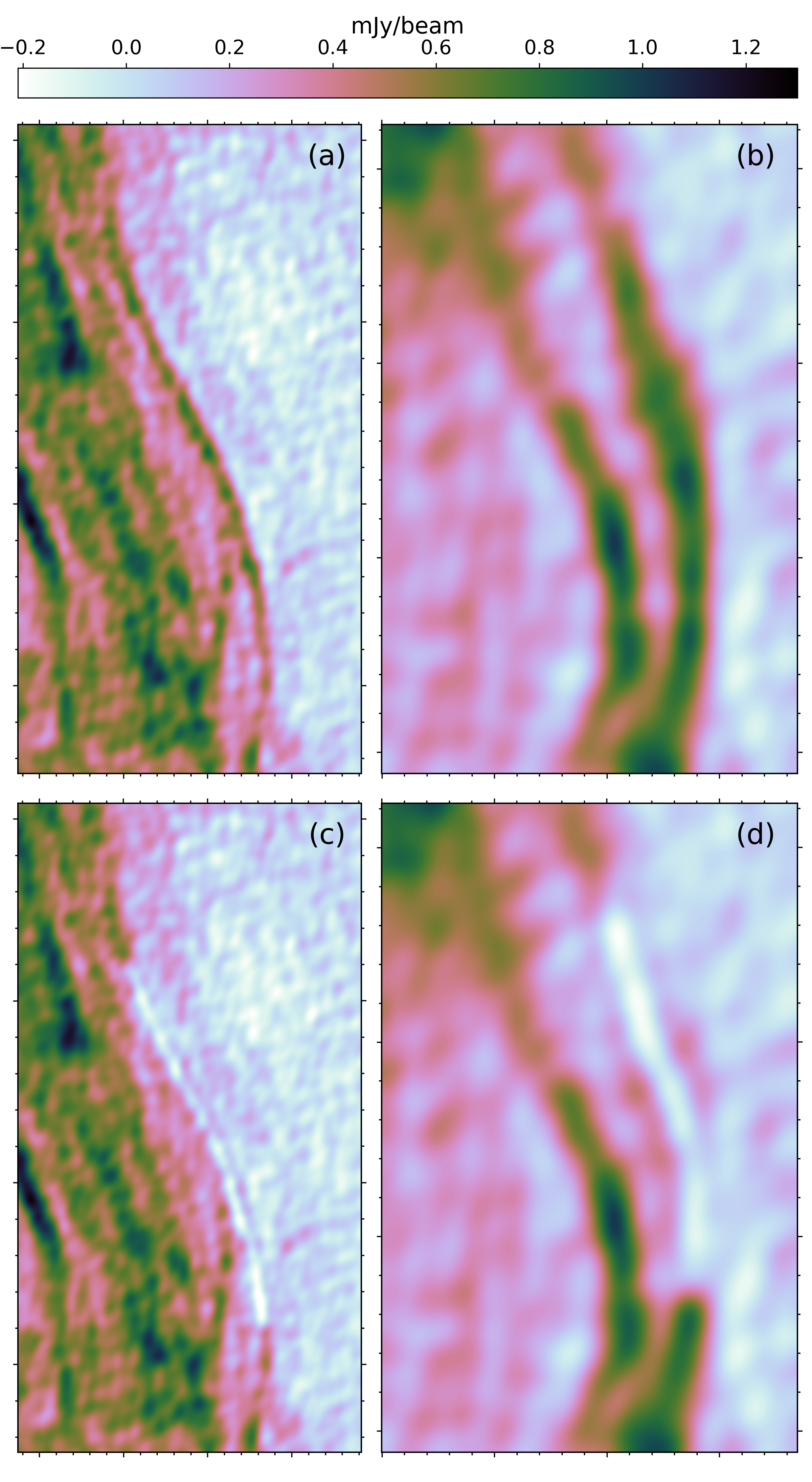}
    \caption{Zoomed-in regions of the LOFAR-VLBI image shown in Fig.~\ref{fig:lofar_map} showing the best \(\chi^2\) fits for the western threads of 3C\,338. The left column (panels \textit{a} and \textit{c}) show the isolated thread in the center of the bundle and the right column (panels \textit{b} and \textit{d}) show the isolated thread at the southern end of the bundle. The top row (panels \textit{a} and \textit{b}) shows the original image and the bottom row (panels \textit{c} and \textit{d}) shows the residuals left after subtracting the best-fitting model for the thread.}
    \label{fig:mcmc_fit}
\end{figure}

The most accurate modelling of the threads requires relatively isolated examples of threads, with the least background emission or interaction with other threads. Therefore, we selected two specific regions: the outer thread in the center of the western threads, and the outer thread at the southern end of the western threads. Both a zoom-in of the original image and the fit residuals are shown in Fig.~\ref{fig:mcmc_fit}.

Considerable systematic uncertainties should be assumed for several reasons: First of all, the clean beam of the LOFAR-VLBI observation has a semi-minor axis spanning around 90~pc at this redshift, forming a soft limit on the scale down to which information can be robustly obtained. Assuming a model of what the source structure is below this resolution, information can still be derived, but only based on the validity of the assumed source model. The probability distribution function for the thread radius we obtain from the MCMC is predominantly located below the 90~pc threshold. Secondly, we only fitted a rudimentary model of a cylindrical shape with a uniform emissivity and radius along its entire length. It is probable that these two parameters in reality will change along the thread, and we also have not accounted for any emission from the environment of the threads. We attempted to add a simple uniform background level to the fit, but did not find that this provided better-fitting results. As can be seen in Fig.~\ref{fig:mcmc_fit}, the best fits provide reasonable models, but the residuals are not indistinguishable from pure noise. Thirdly, while the MCMC fitting from the two separate threads resulted in matching probability distribution functions, it is not guaranteed that the same model parameters would be valid for threads located elsewhere in 3C\,338. For these reasons, we consider that the results from the MCMC fitting only provide a crude indication for the most probable physical parameters.

%%%%%%%%%%%%%%%%%%%%%%%%%%%%%%%%%%%%%%%%%%%%%%%%%%

% Don't change these lines
\bsp	% typesetting comment
\label{lastpage}
\end{document}